# Reflexion on a method for Archaeology of technical machine: a multi-competences experience on roman wooden force pump of Perigueux

N.Perry, Technological Institute of Bordeaux1 University, Fr
R.Stein, University of Reading, UK
R.Vergnieux, Asonius Institut of Bordeaux3 University, Fr
L.Cabot, Wood Technical Institute for Building of Perigueux, Fr

**Abstract**
This study is based on Roman wooden force pumps. It appears that they were used in small numbers to raise water from wells, and more commonly as portable pumps to fight fires. The force pump is attributed to Ctesibius of Alexandria (fl. c.270 B.C.). The earlier examples were made in bronze, but the original design was cleverly re-engineered in Roman times to make pumps easier and cheaper to make and maintain, by cutting apertures in a large block of wood, and making internal spaces pressure proof by plugging their extremities. Eighteen wooden pumps have been found, mostly in wells, and remains of thirteen survive. This study is based on examination of the remains, and of the reports of the Perigueux pump (Dordogne, France).
Archaeologists want to fully study the pump, tacking into consideration its efficiency and the mechanical calculations of the possible performance and of the manpower required to drive it. Technical considerations affecting construction, operation, and performance are discussed and requires connections with knowledge fields such as mechanics. A detailed Report is provided on the pump, giving, so far as possible, a complete description of it and all that is currently known about it. Moreover, a model, size 1:1 of the pump had been built in order to validate the hypothesis. This paper will present the different sides of the multi-expertise for this archaeology and experimental archaeology validation.

## 1. PURPOSE OF THE ARCHEOLOGICAL STUDY

The purpose of this study is to describe and analyse as fully as possible an ingenious Roman machine - the wooden two-cylinder force pump. This machine is important for three principal reasons.

First, it demonstrates that Roman engineers were very capable of taking a Greek machine and transforming it into one which was much easier and cheaper to make than one that followed the original design. This process is very familiar to us today; much of modern production is the result of considering again and again how a design can be improved by altering any and every aspect of it including (for example) materials used, method of manufacture, and ease of maintenance and repair. Well documented examples from the ancient world are not numerous, but in the case of the force pump the change in design is very clear, and very striking. In the original design the pump is made of bronze parts soldered together, but in the new design it is made by cutting apertures into a single large block of oak, and then plugging their extremities to make the resulting vessel proof against high internal water pressure.

Second, the great achievements of Roman civil engineering are evident. Their temples, palaces, fortresses, baths, walls, roads, and aqueducts are very well known and often extensively documented. Many of these structures survive very well because they are built largely of stone or brick. In comparison, the achievements of Roman mechanical engineering are little known and little documented. Comparatively few Roman machines survive; being made largely of rope, wood, leather, and metal, they become victims of recycling and decay.

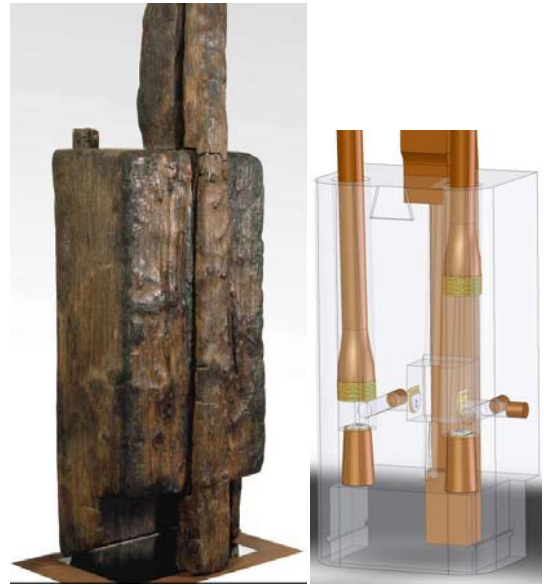

Figure 1: wooden pump body picture (left) and its CAD design

For many types of machine (for example the Archimedes screw; the crane; machines for cutting stone) the number of cases in which substantial remains of the machines themselves have been found is small, and in many cases their existence and configuration can only be surmised from images of them on tombs and other places; from the remains of the emplacements built for them; from the traces that they have left on their surroundings; from objects associated with them; or from the products made by using them. But in the case of the two-cylinder wooden force pump the remains of eighteen actual machines have been found.

Although many were badly decayed, some were (and in some cases still are) in very good condition, and nearly complete. Almost all of the known examples were used in fixed installations (i.e. they were not portable), and this comparatively large set of examples enables us to gain a good understanding of how the machine was configured, and how it functioned, when used in fixed installations.

Finally, the wooden pump gives us an insight into an important means of fighting fire - an ever present and much feared danger - in the ancient world. Many ancient texts refer to pumps being used to fight fires, or to the pumpers who operated them. It is of course possible that some, or many, of the pumps used to fight fires were made of metal; but it seems likely that the cheaper wooden machine was often preferred, and it may well be that the principal use of the wooden pump was not to raise water in fixed installations, but to fight fires. It is probable that of the eighteen known wooden two-cylinder pumps only one (that found in 1908 in the cellar of the Amphitheatre at Trier) was a portable pump used (inter alia) to fight fires, but a study of the wooden pumps that were used in fixed installations enables us to suggest how the portable pump operated, and the performance of which it was capable.

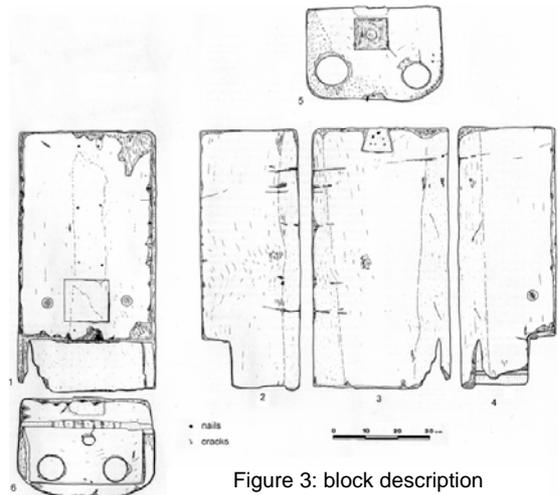

Figure 3: block description

The roman pump is a wooden two cylinder pump. It was found in 1972, in the ruins of the Preigueux's domus, founded close to the Vesone Temple. This pump body was stored in a 10m deep hole, in a mix of water, sand and clay. As you can see in Figure 1, the pump is then very well conserved. His function seems to be dedicated to garden and agricultural use. One of the particularities of this pump is that, Romans describe the need of maintenance. That means the need of extracting the pump from the well; clean the sand and the clay from the valves and cylinders.

The Figure 2 shows the plugs on the front and left faces closing the connecting passages; the sealing plate closing the valve chamber; the filter chamber at the bottom of the block, and the enigmatic bore in its ceiling; the dovetail mortise on the back face; and the square mortise on the top face, with the outlet passage in its centre. The archaeological drawings were the basics document used for 3D CAD redesign, mechanical analysis and model manufacturing. All these points will be quickly presented in the following parts of the paper, focusing on the meets of different knowledge and experiences.

## 2. PROJECT ACTORS

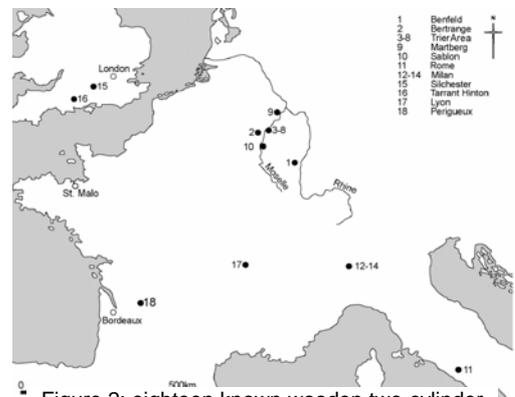

Figure 2: eighteen known wooden two-cylinder pumps positions

We will describe the different actors of this project before pointing out the different contributions without going in details of each contibutor.

First of all, the main actor and contributor for roman wooden force pumps is Dr R.Stein from the School of Human and Environmental Sciences, Department of Archaeology at the University of Reading (1 – 2). He defended his PhD proposal in March 2007, after an analysis of 18 wooden pumps spread all over Western Europe. His work made comparisons between the different pumps, their technical solutions and performances, social use, and he tried to answer the question of the mobility of such tool, form a well to another. The Perigueux's pump is in the Vesone museum of Perigueux, available for deep investigations.

In this context, the Asonius Institute joined the group of actors searching around this pump thanks to M Vergnieux and Mora. This institute (3) is a French national resource for 3D virtual reality platform (called Archeovision) that helps to redesign archaeological and antic objects. It helps archeologist to recreate virtual environments in order to validate, with the virtual immersion, hypotheses. The main well known international studies are the Circus Maximus as illustrated in Figure 4. The Archeovision environment provides the full numerical chain. It allows starting either from small object scans, or big architectures digitalisation, to the full 3D integration and virtual representation.

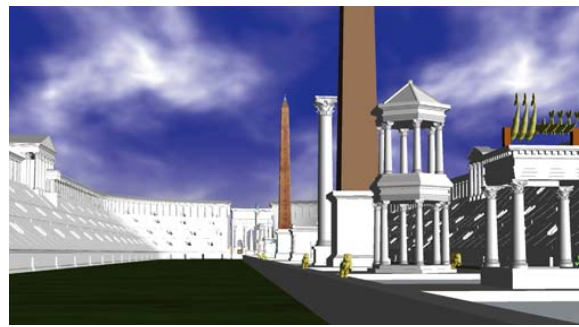

Archeovision made a pump scan using Magnetic Resonance Imaging (M.R.I.) to redraw a virtual/Real wooden pum as presented in figure 5. Moreover, the M.R.I. helps interring within the pump without any physical intrusion. Questions relatively to the presence or not of nails, some accessibility, and the exact measurements of some dimensions could have been done. A major role played by Archeovision, is the project reviewer, using their 3D immersive showroom. Consequently, each actor, can visualise, and comment on the 3D artefacts, pointing out its specific questions or requirements. This show room helps all the different contributors to reach understandings and agreements while seeing the 3D pump representation.

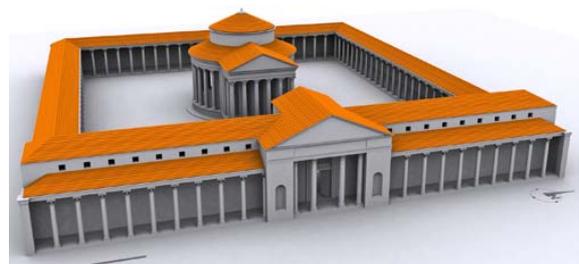

Figure 4: 3D representation of the Circus Maximus (up) and Vesone Temple of Perigueux (down)

The Wood Technical Institut for Buidling of Perigueux helps going from the virtual to the real. Indeed, the teachers of this institute, M Cabot for the history side and M Forget for the wood manufacturing technique aspects, are involved in the Legio VIII Augusta, witch is a group of volunteers that make historical reconstitution (5). They practice the experimental archaeology while living during several days lake legionnaires, walking, eating, sleeping in the same conditions. They can then validate some technical aspects of the real life, real use of devices. The two

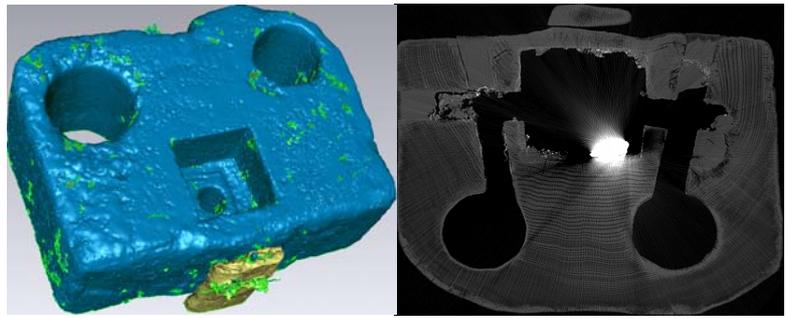

Figure 5: 3D re-design (up) and MRI layer image (down)

teachers made a project with the students in order to manufacture a 1:1 size model of the pump.

This is a 3 years pedagogical project that link historical researches, technical applications of a specific profession, and an applicative project that guide the last year for the students. It took a full year to analyse, draw in detail each part of the pump, find the solutions for the wood remanufacturing, and fabricate the model of the pump (as presented in figure 6). This pump has Plexiglas parts to show the movements of the water and the specific solutions for the valves. This model validate the feasibility and usability of such a pump, and was able to evaluate the coherency of the water flow calculus versus the experimental application.

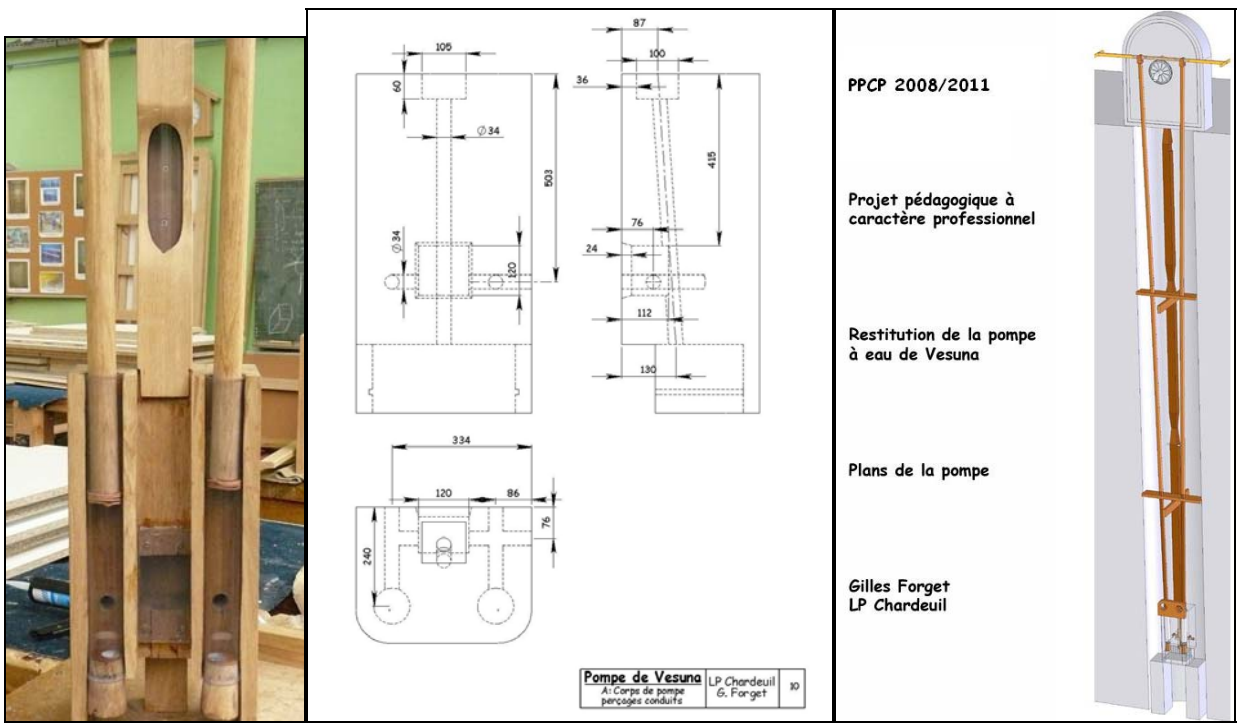

Figure 6: Model of the pump (left), detailed body drawing (center), illustration of the 3 years pedagogical project (right)

The last actors are the students from the Science and Technique Bordeaux1 University. Two levels of students were involved in the project, Master1 degree students and 1rst year at the university students. Their background is mechanics and technology. The objectives of the students are to validate mechanical calculus such as the forces, the flows, the cinematic and mobility of different elements. Based on their analysis, the Master students tried to formalise a dedicated method to face problems coming from the archaeologist's field and requiring engineering approaches. We will quickly present the method in the next part.

## 3. FIRST ELEMENTARY BRICKS FOR THE METHODOLOGY PROPOSAL

The methodological proposal is based on a one side of literacy analysis (6, 7, 8, 9) on the possibilities for old technical machines to have a second life (real or virtual) in museum environments. And, on the other side, from the distance and look back we made on this project.

Their is no revolution in our proposal, but it is more an identification of what we think relevant in term of successive steps. Based on the system analysis, we propose to analyse the technical system with three directions: functional, physical and modelling analyses.

### 1. Functional analysis

Based on the functional analysis approach, we propose to identify, what ever may be the exact design of the product, to identify the core functions of the device. To enrich this study, all the different phases of use hove the product have to be taken into account to check all the needed functionalities. For example, in the case of the wooden force pump, Romans used to make maintenance and cleaning of the pump, witch means that the pumps have to be easily removable and handle from the well. Figure 7 presents a result of the analysis like on the top a functional diagram and bellow functions definition table. These functions can be specified in term of expected value, if you design a new product, or become the data that have to be validated or evaluated for the archaeological point of view.

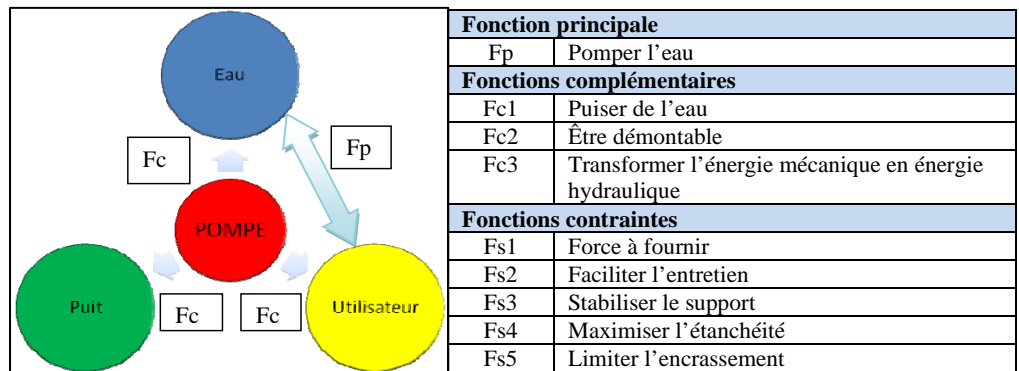

| **Fonction principale** | |
|---|---|
| Fp | Pomper l'eau |
| **Fonctions complémentaires** | |
| Fc1 | Puiser de l'eau |
| Fc2 | Être démontable |
| Fc3 | Transformer l'énergie mécanique en énergie hydraulique |
| **Fonctions contraintes** | |
| Fs1 | Force à fournir |
| Fs2 | Faciliter l'entretien |
| Fs3 | Stabiliser le support |
| Fs4 | Maximiser l'étanchéité |
| Fs5 | Limiter l'encrassement |

Figure 7: functional analysis result

## 2. Physical analysis

The second step is the analysis from the existing elements point of view. Luckily, for the pump, most of the compounds were founded, in a good estate. The puzzle was not so hard to rebuilt, even if many different architecture could exist for a double cylinder pump. A part list tree is then built, making the link between the architectural definition of the product and the extracted and identified elements from the archaeological work site.

The FAST (Functional Analysis System Technic) diagram introduce the different components going from the function to the solution (see Figure 8). In addition, a schematic representation of the system is proposed (see figure 9) that propose the architectural definition of the product. This schema is the first base for identification of the physical parameters that will be necessary for all the further calculation and evaluations.

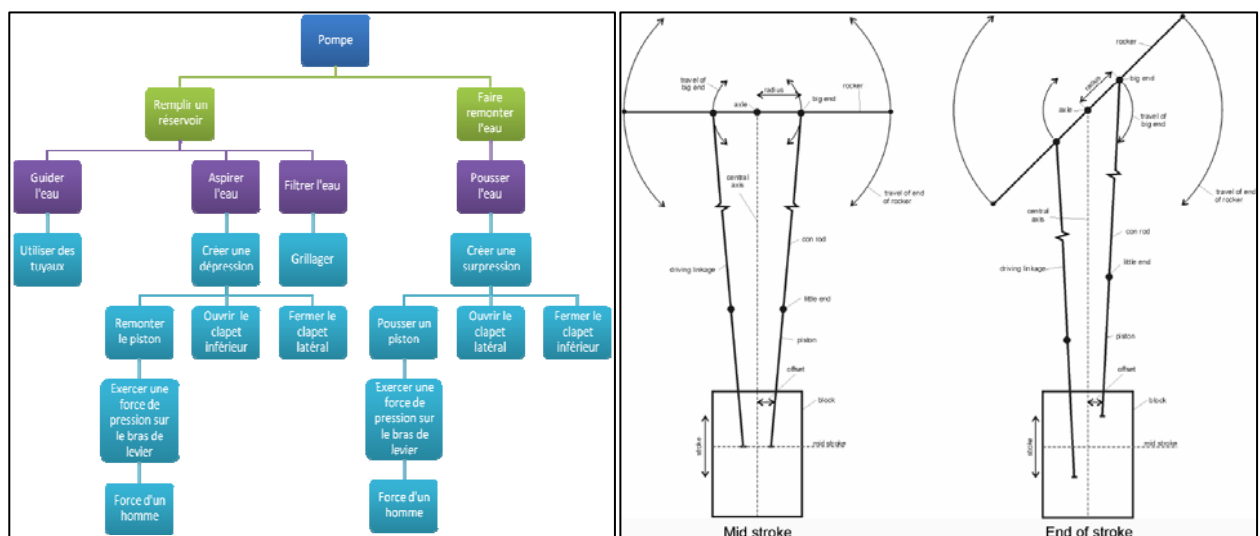

Figure 8: From Function to component (left) and architectural schema (right)

At the end of this phase, the archaeologist can specify the specific information available, and the need of deeper definition or calculation to validate hypotheses.

## 3. Models analysis

The two first steps defined thanks to the functional analysis the expected functionalities, and thanks to the physical analysis the needed data for the evaluation of the functions. It is now necessary to identify what physical models represent the behaviour of the machine. Taking into account the limited needs of accuracy, but much lore a good evaluation, a selection of models have to be done. In a first stage, priorities to macro-models, more global are the first introduction to answers about the technical questions of the archaeologist. In a second time, if specific observations or incoherency between the hypothesis and the results are identified, more precise models will be needed.

In the case of the force pump, we used the conservatives' equations: energies and mass conservation to evaluate the efforts needed for making the water flowing. More we take the quasi static hypothesis in order to simplify the effects of inertias. Finally we neglected all the pressure loss in the 10m tube from the pump body to the top.

For the cinematic evaluation, we used both an analytical evaluation and a CAD simulation of the different movements.

The models used were only the fundamental principle of the static, the energy conservation, the hydrostatic law. In addition we evaluated a water loss ration and a friction loss that were deducted to the results. We take specific attentions to the ergonomics impacts of some hypothesis, taking into account that even if the final use is not well defined, a long period uses need not to big movements' amplitudes, a reasonable frequency and no extraordinary efforts.

It still remains a full CAD simulation, linking the cylinders position to the opening or closing of the valves, and simulating the water flow.

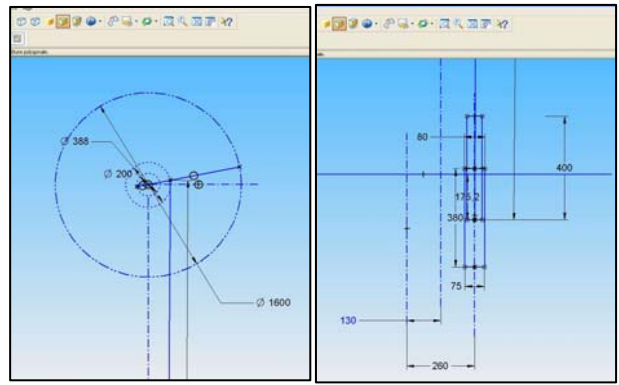

Figure 9: Cad simulation of the movements

## 4. CONCLUSION

The necessity of a multi competence approach is obvious if a historical technical system is studied. But the different communities have to learn to work together. Cultural, background, vocabulary and scientific approaches differences dig gaps between the experts. If the actors reach to bridge the gap, the gains for the projects are important.

Based on these elements, we try to propose a systematic method that will guide both the archaeologists to clarify their requirements and express clearly the expected results. But also a method, adapted from the functional analysis, that present all the basics elements needed to face this kind of problem, proposing tools solution integrated in the method and easily acceptable by the archaeologists.

A second project on another kind of water machine, from Barzan thermal spring (Charente Maritime, Fr) will be tested with this method in order to enrich the systematic demarche.

## 5. AKNOWLEDGMENTS

The authors would like to thanks R.Stein for his huge amount of work on the wooden force pump, the Archeovision technical support, for all the scanning and virtual rebuilding of the pump, the students and their teachers from the Wood Technical Institute for Building of Perigueux, that gave the possibility to manufacture a 1:1 model and test it, and the Technological Institute of Bordeaux1 University and their teachers that found funny support to apply mechanical knowledge, far from aeronautic or automotive applications. But the main thanks rely to the Roman and more generally antics intelligence and technological addaptativity.